# Dark Matter Universal Properties in Galaxies

Christiane Frigerio Martins

*Universidade Federal do ABC, Rua Catequese 242, 09090-400 Santo André-SP, Brazil*

**Abstract.** In the past years a wealth of observations has unraveled the structural properties of dark and luminous mass distribution in galaxies, a benchmark for understanding dark matter and the process of galaxy formation. The study of the kinematics of over thousand spirals has evidenced a dark-luminous matter coupling and the presence of a series of scaling laws, pictured by the Universal Rotation Curve paradigm, an intriguing observational scenario not easily explained by present theories of galaxy formation.

**Keywords:** dark matter - galaxies: kinematics and dynamics.
**PACS:** 98.62.-g, 95.35.+d

## INTRODUCTION

It has been known for several decades that the kinematics of disk galaxies exhibit a mass discrepancy [1, 2, 3]: spirals show an inner baryon dominance region [4, 5, 6] whose size ranges between 1 and 3 disk exponential length-scales according to the galaxy luminosity [7], inside which the observed ordinary baryonic matter accounts for the rotation curve (RC), but outside which, the distribution of the baryonic components cannot justify the observed profiles and sometimes the amplitudes of the measured circular velocities [8, 9, 10]. This is usually solved by adding an extra mass component, the dark matter (DM) halo. RCs have been used to assess the existance, the amount and the distribution of this dark component.

In Spirals we have a unique opportunity to investigate their global mass distribution: the gravitational potentials of a spherical stellar bulge, a dark halo, a stellar disk and a gaseous disk give rise to an observed equilibrium circular velocity

$$V^2_{tot}(r) = r \frac{d}{dr} \phi_{tot} = V^2_b + V^2_{DM} + V^2_* + V^2_{HI}.$$

The Poisson equation relates the surface (spatial) densities of these components to the corresponding gravitational potentials. It is not difficult to estimate their contributions. For example the surface disk stellar density is proportional (by the mass-to-light ratio) to the observed Freeman surface brightness [11], $\Sigma_*(r) \equiv M_D/(2\pi R_D^2) e^{-r/R_D}$, where $M_D$ is the disk mass and $R_D$ is the disk length-scale[1], the latter being measured directly from the observations. Then $V^2_* \equiv G M_D/(2 R_D) x^2 B(x/2)$, where $x \equiv r/R_D$, G is the gravitational constant and the quantity $B \equiv I_0 K_0 - I_1 K_1$ is a combination of Bessel functions. The surface gaseous density is directly obtained by HI measurements.

---

[1] It is useful to define the optical radius (the radius enclosing 83% of the total light [12]), $R_{opt} = 3.2\, R_D$, as the "size" of the stellar disk.

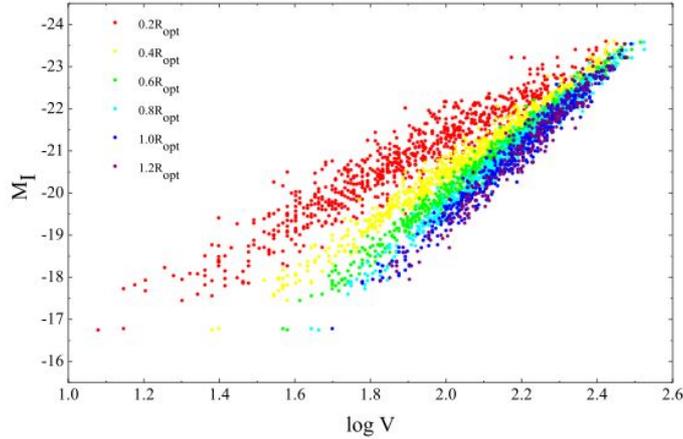

**FIGURE 1.** The Radial TF. The relations at different radii are indicated with different colours [14].

The assumption that the RC of a spiral leads to a fair measure of its underlying gravitational potential is well justified: i) in their very inner regions the light well traces the gravitational mass [13] ii) there exists, at any galactocentric radii measured in terms of disk length-scale $R_n \equiv (n/5)R_{opt}$, a *radial* Tully-Fisher relation [14] linking with very low scatter the local rotational velocity $V_n = V_{rot}(R_n)$ with the total galaxy luminosity (see Fig.1) $M_{band} = a_n log V_n + b_n$ ($a_n$ and $B_n$ are the slope and zero-point of the relations) iii) non-circular motions are only present at the level of a few km s$^{-1}$ (and are not associated with a global elongation of the potential), as found from harmonic decomposition by [15] and by [16] of the velocity field of many galaxies from the HI Nearby Galaxy Survey [17] iv) disk mass estimates from spectro-photometric and kinematical methods are statistically equivalent (see e.g. [18]).

Contrary to what is often claimed, the observational evidence indicates that RCs are not asymptotically flat [19, 20] neither flat inside $R_{opt}$ [21]. When in the late 1970s the phenomenon of DM was discovered a few truly flat RCs were highlighted in order to rule out the claim that non Keplerian velocity profiles originate from a faint baryonic component distributed at large radii. At that time a large part of the evidence for DM was provided by extended, low-resolution HI RCs of very luminous spirals whose velocity profile did show small radial variations.

The increase in the quality of the RCs though soon leads to the conclusion that baryonic (dark) matter was not a plausible candidate for the cosmological DM and that the RCs did show variation with radius, even at large radii. Later numerical simulations in the Cold Dark Matter (CDM) scenario also predicted asymptotically declining RCs [22]. The flat RC paradigm was hence dismissed in the 1990s (e.g. [5, 23, 24]). Today, the DM halo structure and its rotation speed has a central role in Cosmology and is thought to have a strong link to Elementary Particles via the Nature of their constituents (e.g. [25]). A careful interpretation of the spiral RCs is then crucial.

Note that the circular velocity due to a Freeman stellar disk has a flattish profile between 2 and 3 disk length-scales implying that a flat RC is not necessarily a proof for the existence of DM. Its most solid evidence instead originates from the fact that even in very faint galaxies the RCs are often steeply rising already in their optical regions. Fig. 2 shows, for a large sample of galaxies, the logarithmic slope $\nabla$ of the circular velocity at $R_{opt}$ as a function of $M_B$ and $V_{opt}$: it takes almost all the values allowed by Newtonian gravity, from -0.5 (Keplerian regime) to 1 (solid body regime) and, furthermore, it strongly correlates with galaxy luminosity and $V_{opt}$ [12, 19, 26, 27].

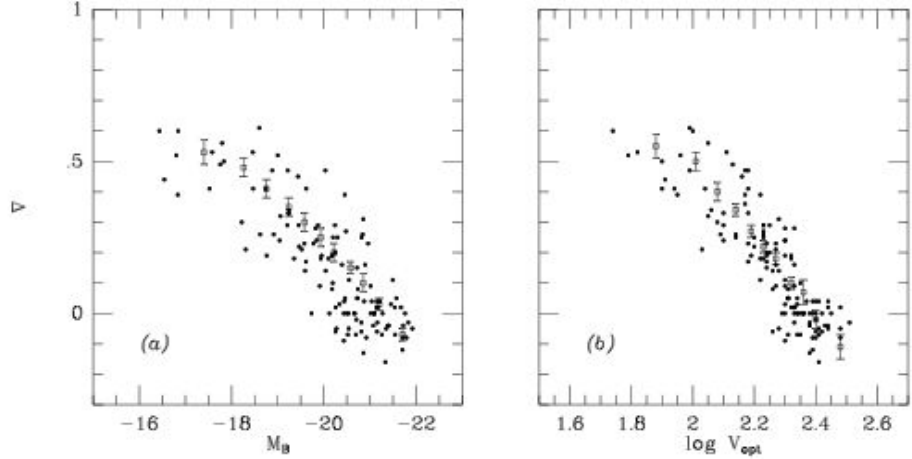

**FIGURE 2**. The RC log slope as a function of $M_B$ and $V_{opt}$ of 1100 coadded and 100 individual objects [12].

## DARK HALOS FROM OBSERVATIONS: THE UNIVERSAL ROTATION CURVE

The study of the systematics of spiral kinematics, pioneered by Persic & Salucci 1991, further developed by Persic, Salucci & Stel 1996 and by Salucci *et al.* 2007 has evidenced that these systems present universal features well correlating with global galactic properties. This has led to the construction of the "Universal Rotation Curve", $V_{URC}(r; P)$, i.e. an empirical function of galactocentric radius r, that, tuned by a global galaxy property (e.g. luminosity), is able to reproduce the RC of any object[2] (see Fig. 3). Additional kinematical data and virial velocities [28], $V_{vir}=(GM_{vir}/R_{vir})^{1/2}$, have determined the URC out to the virial radii.

$V_{URC}$ is then the observational counterpart of the velocity profile that emerges out of CDM large N-body numerical simulations of structure formation. As individual RCs, it implies a mass model including a Freeman disk and a DM halo with an empirical Burkert cored profile [29],

$$\rho(R)=\frac{\rho_0 r_0^3}{(R+r_0)(R^2+r_0^2)},$$

where $r_0$ is the core radius and $\rho_0$ its central density.

---

2 A 3D visualization can be found at http://www.youtube.com/watch?v=YcgafVb-WJI.

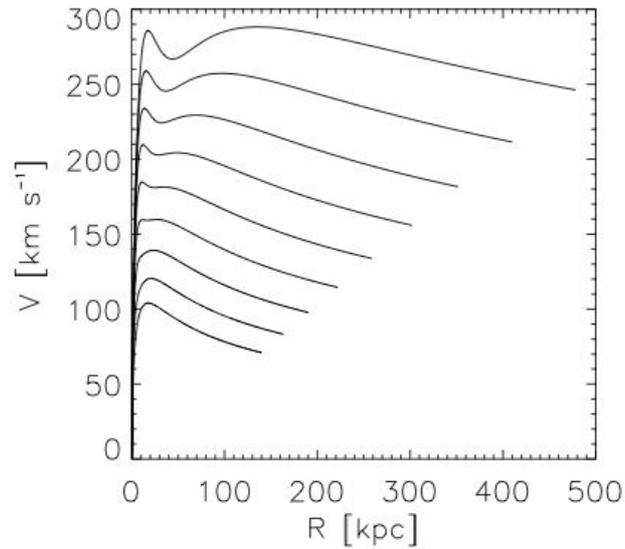

**FIGURE 3**. The URC [20]. Each curve corresponds to a galaxy mass $M_{vir}=10^{11}10^{n/5}M_\odot$, with n=1...9 from the lowest to the highest curve.

Recent debate in the literature has focused on the "cuspiness" of the DM density distribution in the centres of galaxy halos predicted from the simulations [22, 30, 31, 32], commonly represented by the Navarro, Frenk & White (NFW) profile, but not seen in observed data, as well as in the various systematics of the DM distribution (e.g. [9, 15, 20, 33, 34, 35, 36]).

It is worth illustrating one example of this disagreement: the nearby spiral dwarf galaxy DDO 47 [15]. The RC mass modeling (Fig. 4) finds the dark halo with a core radius of about 7 kpc and a central density of $1.4\times10^{-24}$g cm$^{-3}$: the underlying DM density profile is much shallower than the NFW predictions, totally unable to fit the RC. Presently there are about a hundred high quality, extended and free from deviations from axial symmetry RCs whose careful analysis has strongly disfavored the disk + NFW halo mass model (in favor of cored profiles) that: i) fits the RC poorly ii) implies an implausibly low stellar mass-to-light ratio and in some case iii) an unphysical high halo mass (e.g. [9, 15, 36, 37, 38, 39, 40, 41]). Furthermore, these evidences are accompanied by investigations that have ruled out that it may arise from (neglected) systematical effects [15, 16, 42] (for a recent review on the cusp versus core issue see [43]).

The structural parameters $\rho_0$, $r_0$ and $M_D$ are obtained for the URC and for any individual RC by $\chi^2$ fitting. As result, a clear scenario of the mass distribution emerges[3], with a cored DM distribution and a set of scaling laws among local and global galaxy quantities:

- Spirals have an inner baryon dominance region where the stellar disk dominates the total gravitational potential, while the DM halo emerges farther out
- At any radii, objects with lower luminosities have a larger dark-to-stellar mass ratio.

---
3 A worldwide initiative specially devised to increase awareness of the phenomenology of DM in galaxies can be found in http://darkmatteringalaxies.selfip.org.

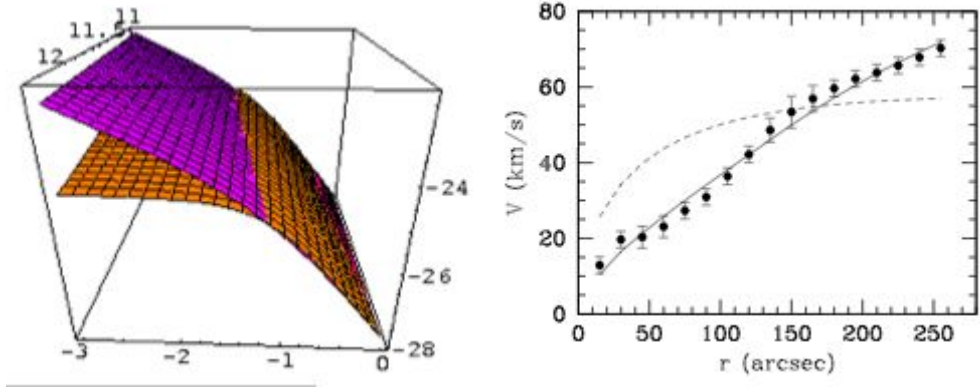

**FIGURE 4.** Left: The URC halo density *vs* the NFW halo density of objects of the same mass, as a function of normalized radius and virial mass [20]. The axes labels are $R/R_{vir}$, $\log M_{vir}/M_\odot$ and $\log(\rho(\text{g cm}^{-3}))$. Right: RCs best-fits of DDO 47 [15]: Burkert halo + stellar disk (solid line), NFW halo + stellar disk (dashed line).

The baryonic fraction in spirals is always much smaller than the cosmological value $\Omega_b/\Omega_{matter} \simeq 1/6$, and it ranges between $(7-50) \times 10^{-3}$, suggesting that processes such as SN explosions must have removed a large fraction of the original hydrogen
- Smaller spirals are denser, with their central density spanning 2 orders of magnitude over the mass sequence of spirals
- The stellar mass-to-light ratio (in the B band) lies between 0.5 and 4 and increases
- with galaxy luminosity as $L_B^{0.2}$, in agreement with the values obtained by fitting their SED with spectro-photometric models
- Dark and luminous matter are remarkably linked together (see Fig. 5).

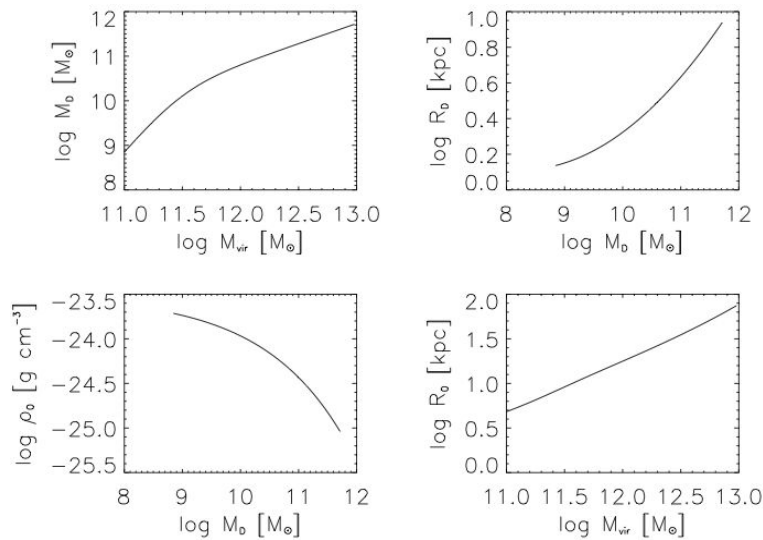

**FIGURE 5.** Scaling relations between the structural parameters of the dark and luminous mass distribution in spirals [20].

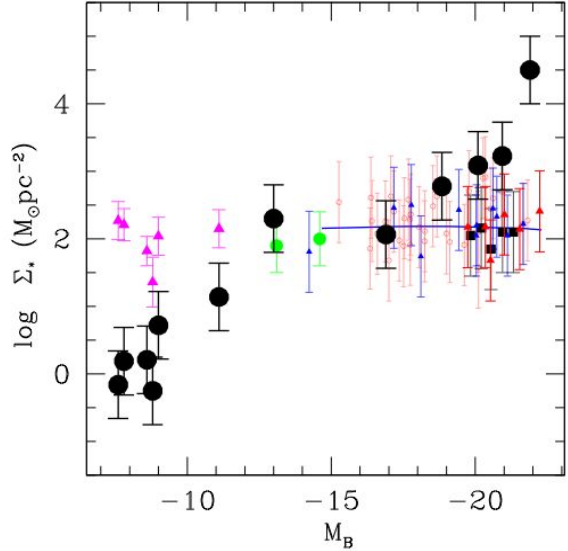

**FIGURE 6.** Dark matter central surface density in units of $M_\odot pc^{-2}$ as a function of galaxy magnitude, for different galaxies and Hubble Types [44]. As a comparison the values of the same quantity of the stellar component is also shown (big filled circles).

## A CONSTANT DARK MATTER HALO SURFACE DENSITY IN GALAXIES

An intriguing finding among the DM structural properties is that the central surface density, $\mu_{0D} \equiv r_0 \rho_0$, is nearly constant and independent of galaxy luminosity. Based on the co-added RCs of ~1000 spirals, mass models of individual dwarf irregular and spiral galaxies of late and early types with high-quality RCs, and galaxy-galaxy weak lensing signals from a sample of spirals and ellipticals, we find that [44]

$$\mu_{0D} = 2.15 \pm 0.2,$$

in units of $\log(M_\odot pc^{-2})$. We also show that the observed internal kinematics of Local Group dwarf spheroidal galaxies are consistent with this value. Our results are obtained for galactic systems spanning over 14 magnitudes, belonging to different Hubble Types, and whose mass profiles have been determined by several independent methods.

The approximate constancy of $\mu_{0D}$ is particularly relevant also because it is in sharp contrast to the observed systematical variations, by several orders of magnitude, of the central stellar surface density, i.e. of its luminous counterpart, as illustrated in Fig. 6. As a consequence, the central surface density is the only DM quantity which is not correlated with its stellar analagous, different from any other (core radius, central spatial density, mass, etc). Moreover it is remarkable that the constancy of $\mu_{0D}$ can be related to the above scaling laws of spirals. As an example, let us define $M_{h0}$ and $V_{h0}$ as the enclosed halo mass inside $r_0$ and the halo circular velocity at $r_0$. Since for a Burkert halo $M_{h0} \propto \rho_0 r_0^3$, then $M_{h0} \propto V_{h0}^4$, which immediately reminds a sort of Tully-Fisher relation.

It is surprising and presently difficult to understand how the DM halo central surface density can be maintained constant across galaxies with very different evolutionary histories (e.g. numbers of mergers, significance of baryon cooling, stellar feedback) and with range from DM-dominated to baryon-dominated in the inner regions. This quantity may hide an important physical meaning in the nature itself of the DM, as it seems to indicate recent theoretical work [45, 46, 47].

## CONCLUSIONS

More then eighty years ago E. Hubble established the expansion of the Universe with his pioneering observations of galaxies. Since then galaxies have been fundamental tools for understanding the structure and evolution of our Universe. Today they are crucial laboratories where microphysics phenomena, up to now not detected by particle physics experiments, emerge with unprecedented clarity.

The distribution of luminous and dark matter in galaxies shows amazing properties and a remarkable systematics that make it as one of the hottest cosmological issues. There is no doubt that this emerging observational scenario will be decisive in guiding how the CDM-based theory of galaxy formation must evolve to meet the challenge that the observational data are posing.

## ACKONWLEDGMENTS